\documentclass{revtex4}

\usepackage{graphicx}

\usepackage{amssymb}

\begin{document}

%\begin{frontmatter}

\title{Structure, stability, and stress properties of amorphous
and nanostructured carbon films}

\author{M.G. Fyta$^1$, C. Mathioudakis$^2$, G. Kopidakis$^2$, and P.C. Kelires$^1$}
\affiliation{$^1$Physics Department, University of Crete, P.O. Box 2208, 
 710 03, Heraclion, Crete, Greece, \\
$^2$Department of Materials Science, University of Crete, 710 03 Heraclion, Crete, Greece.}

\begin{abstract}
Structural and mechanical properties of amorphous and nanocomposite
carbon are investigated using tight-binding molecular dynamics
and Monte Carlo simulations. In the case of amorphous carbon, we show 
that the variation of $sp^3$ fraction as a function of density is linear
over the whole range of possible densities, and that the bulk moduli follow
closely the power-law variation suggested by Thorpe. We also review
earlier work pertained to the intrinsic stress state of tetrahedral
amorphous carbon. In the case of nanocomposites, we show that the
diamond inclusions are stable only in dense amorphous tetrahedral
matrices. Their hardness is considerably higher than
that of pure amorphous carbon films. Fully relaxed diamond 
nanocomposites possess zero average intrinsic stress.

\end{abstract}
\maketitle

\section{INTRODUCTION}
\label{sec:intro}

Amorphous carbon (a-C) has been intensively investigated over the
years, but its properties are not fully understood. 
Some of them are strongly debated. Tetrahedral a-C (ta-C), containing a 
high fraction of $sp^3$ hybrids, is the form of a-C which
has drawn most attention because of its
diamondlike properties \cite{Robertson02,Silva02}, including high
hardness for mechanical purposes, a wide band gap for optical 
applications, and biocompatibility for biomedical coatings. 
Ta-C has also promising applications in micro-electromechanical 
devices (MEMS).

Recently, nanostructured amorphous carbon (na-C) has attracted 
attention \cite{Sabra98,Banhart99}. It is a hybrid form of carbon in which 
nanocrystallites are embedded in the a-C matrix. These range from 
nanodiamonds \cite{Lifshitz}, to open graphene structures 
with negative curvature (schwarzites) \cite{Vanderbilt,Donadio},
to carbyne films (composed of $sp^1$ chainlike structures) 
\cite{Barborini,Ravagnan}. The variety of embedded nanostructures, 
and the great number of possible configurations in the matrix, open
some new ways to explore the physics, tailor the mechanical and electronic 
properties, and extend the applications of a-C.

From a fundamental point of view, the most important characteristic
of na-C is its inhomogeneous nature, characterized by large gradients of 
density and coordination through the system. The challenge for the theorist,
therefore, is to provide a global description of the composite
material, extending and/or generalizing concepts and trends which
apply to the better understood single-phase system. 

Here, we present recent work \cite{Fyta03,Mathiou04} aiming at such
a theoretical description. It is based on tight-binding molecular
dynamics (TBMD) and Monte Carlo (MC) simulations. We first review
work on single-phase a-C pertained to its structure, stress
state, and to physical trends followed by the $sp^3$ fraction and
elastic moduli as a function of density and mean coordination. We then
discuss na-C, focusing on diamond nanocomposite films. Emphasis is given 
on their structure, stability, stress state, and hardness. 
One of the important findings is that the hardness of nanocomposite 
films is considerably higher than that of single-phase a-C films.

\section{METHODOLOGY}
\label{sec:method}

In much of the work presented here, we treat the interatomic 
interactions in a-C networks using the tight-binding (TB) method. 
This bridges the gap between classical and first-principles 
calculations. It is more accurate and transferable
than empirical schemes, providing a quantum-mechanical 
description of the interactions, and it yields greater
statistical precision than {\it ab initio} methods, allowing the use 
of larger cells.

We use the environment-dependent tight-binding
(EDTB) model of Tang, Wang, Chan, and Ho \cite{TWCH}. This model
goes beyond the traditional two-center approximation and allows the
TB parameters to change according to the bonding environment. In this
respect, it is a considerable improvement over the
previous two-center model of Xu, Wang, Chan, and Ho \cite{XWCH}. Both
accuracy and transferability are improved, as shown from recent
successful applications \cite{Galli1}.

The TBMD simulations are carried out in the canonical ($N,V,T$) ensemble.
$T$ is controlled {\it via} a stochastic temperature control 
algorithm. The a-C networks are generated by quenching from
the melt. Although not directly related to the kinetics of the growth 
process of a-C films, it produces generic structures associated with 
the equilibrium state of the films. (See relevant discussion 
in Ref. \cite{Kel00}.)
Cubic computational cells of 216 and 512 atoms with periodic boundary 
conditions are used. Quenching at different durations and rates 
was performed to check the effect on the properties. 
The longer run was for 52 ps and the rate was 226 K/ps. 
Two other runs lasted for 26 and 12 ps, at 226 K/ps and 500 K/ps, 
respectively. No significant changes were found.
During quenching, the volume/density of the 
cells was kept constant. After quenching, the density was
allowed to relax by changing homogeneously the dimensions of the cells
within small increments and seeking energy minimization. The minimum 
energy and the corresponding density and bulk modulus of each cell were 
determined by fitting the energy-versus-volume data to Murnaghan's 
equation of state \cite{Murnaghan}.

Diamond nanocomposite cells are generated by melting and subsequent
quenching a diamond structure, while keeping a certain number of atoms
in the central portion of the cell frozen in their ideal crystal 
positions. After quenching, which produces amorphization of the
surrounding matrix, the cells are thoroughly relaxed with respect to
atom positions and density. Cells with varying coordination (density) 
of the amorphous matrix can be formed by changing the initial starting 
density (volume) of the diamond structure. The size (radius) of the 
nanocrystals is controlled by the choice of the number of the shells kept 
frozen during quenching. Their shape is spherical. The nanocomposite
cells produced by TBMD contain 512 atoms.

The TBMD simulations are supplemented by MC simulations using larger
cells of 4096 atoms to make certain investigations, such as the
stability of nanocomposite films as a function of the density
of the amorphous matrix and the stress analysis, tractable. The
Tersoff potential is used \cite{Terspot}. 

In addition, we examine the properties of the WWW generic
model \cite{Wooten,Djord1}. This is a hypothetical model of 
``amorphous diamond'', completely tetrahedral,
constructed from the diamond lattice by a bond-switching mechanism.
We relaxed its topology and density with the
EDTB model. The WWW model, although hypothetical, is very useful 
because it provides an upper bound to the density, $sp^3$ fraction, 
and bulk modulus of single-phase a-C. Its properties can then 
be compared to the respective ones from diamond nanocomposites.

\section{RESULTS AND DISCUSSION}

\subsection{Single-phase a-C}

We first discuss the pure a-C phase, reviewing past and recent
work. To examine its microstructure, let us look at two 
representative examples, among the sequence of structures characterized
by different density and $sp^3$ fraction. One for the ta-C dense phase, 
and the other for a low-density phase. Their networks are portrayed in 
Fig. 1. 

The ta-C network in panel (a) has a density of 2.99
gcm$^{-3}$, it shows a clear predominance of $sp^3$ bonding (79\%), 
and it reveals that the $sp^2$ sites are largely 
clustered \cite{Frau93,Drabold,Marks96}. Clustering is present in the
form of olefinic, chainlike geometries. The $sp^2$ chains are isolated
and do not link (percolate) to a single spanning cluster, in agreement
with {\it ab initio} work \cite{Marks96}. The driving force behind 
the clustering effect is stress relief. Earlier work \cite{Kel00},
addressing the issue of local rigidity in ta-C, showed 
that clustering contributes stress relief and rigidity to the 
network. This depends on the degree of clustering. The larger the 
cluster, the higher the stress relaxation and the contribution to 
rigidity in the network.

The low-density network in panel (b) has a density of 1.20 gcm$^{-3}$
and contains only 1\% of $sp^3$ sites. It has an open structure with 
long chains and large rings, and with numerous $sp^1$ sites (33\%). 
This network should be typical of cluster-assembled carbon 
films \cite{Barborini,Ravagnan} with an amorphous $sp^2$ character 
and a sizeable carbyne ($sp^1$ chains) component. Such films have
attracted attention for various applications, including field emission, 
catalysis, and gas absorption. 

Analysis of the ring statistics in the ta-C networks reveals the
existence of three- and four-membered rings. (The shortest-path criterion 
of Franzblau \cite{Franzblau} was used to define the ring sizes.)
This is the first tight-binding model which predicts three-membered
rings in ta-C, in agreement with {\it ab initio} MD 
simulations \cite{Marks96} using the Car-Parrinello method.
The five-membered rings are slightly more numerous than the six-membered
rings and significantly more numerous than the seven-membered ones.

The issue of intrinsic stress and its association to $sp^3$ bonding
in ta-C films has been strongly debated over the
years. We have now reached at a rather clear picture of this issue.
We summarize here the important points. Work by McKenzie and 
co-workers \cite{McKenz1} proposed that the compressive stress in ta-C
is produced by the energy of ion bombardment in the deposition process 
which gives rise to local compression, accompanied by the shallow 
implantation of incoming atoms. Their model considers the compressive 
stress as the causative factor for the formation of sp$^{3}$ sites 
and supports the idea of a transition from an sp$^{2}$-rich to an 
sp$^{3}$-rich phase at a critical value of the average compressive stress 
(about 4-5 GPa) which stabilizes the sp$^{3}$ bonding.

While this scenario can not be ruled out for as-grown films, it fails to
describe the stress state of post-growth annealed/relaxed films. It has
now become apparent, after a series of theoretical works by 
Kelires \cite{Kel00,Kel94,Kel01Phy} and thermal annealing experiments
\cite{Friedmann,Ferrari,Kalish,Alam}, that the intrinsic stress is
not a crucial factor for the stabilization of sp$^{3}$ bonding.
A critical {\it average} compressive stress necessary to sustain a high
fraction of sp$^{3}$ sites, as required by the McKenzie model,
does not exist. This conclusion is borne out of the {\it local atomic stress}
model of Kelires \cite{Kel94}, which proposes that the average intrinsic
stress of relaxed ta-C films can be zero, while stress at the atomic level
can be finite and substantial. It further says that the favored stress 
state of sp$^{3}$ sites is compression, while that of sp$^{2}$ sites
is tension, the latter playing the role of relieving stress in the
network. 

According to the local stress model, the as-grown, highly strained and
sp$^{3}$-rich ta-C films are in a metastable state with respect to the 
relaxed, stress-free and still sp$^{3}$-rich ``quasi-equilibrium'' 
ta-C structures. (True equilibrium structures are the 
graphite-like sp$^{2}$-rich films.) The as-grown films
possess high intrinsic stress because the stressed non-equilibrium local
structures are frozen-in during deposition, but the network at the
low deposition temperatures does not acquire enough energy, or it is
very slow at typical times in the laboratory, in order to overcome
the potential barrier between the two states and relax the excessive
stress. Post-growth thermal annealing at moderate $T$'s has proved to 
be a very efficient mechanism for providing the necessary
energy in ta-C films to reach their quasi-equilibrium, stress-free state.
The stress relief can be achieved with minimal structural 
modifications \cite{Ferrari,Sullivan}, without reducing the sp$^{3}$
fraction. However, further annealing above $\sim$ 1200 K transforms 
ta-C into the graphite-like sp$^{2}$-rich phase.

Another issue which is still unclear regards the variation of $sp^3$ 
fraction or, equivalently, of mean coordination $\bar{z}$, 
with density. The basic question underlying this issue is whether there 
is a linear relationship between these two quantities.
We have recently carried out an extensive investigation \cite{Mathiou04}
of this issue through the entire range of densities relevant to a-C,
using the TBMD method. Several networks have been generated, at
various quenching rates, providing sufficient statistics to reach
a definite conclusion. We also compare to the WWW network relaxed with 
the EDTB model.

The variation of $sp^3$ fraction with density is shown in Fig. 2. 
(Hybrid fractions are extracted by counting neighbors
within and up to the first minimum of the pair distribution 
functions, not shown.) Without any doubt, the variation
is linear through the entire range of possible densities.
A linear fit to the points gives
\begin{equation}
\rho (\rm{g/cm}^{3}) = 1.27 + 2.08\ (sp^{3} \rm{fraction}).
\label{den-sp3}
\end{equation}
Eq. (\ref{den-sp3}) predicts the minimum density required to
sustain $sp^3$ bonding in a-C to be $\sim$ 1.3 gcm$^{-3}$. 
The $sp^3$ sites are needed in such low-density networks as linking 
geometries between the main $sp^2$ and $sp^1$ components. For 100\%
$sp^3$ bonding, the corresponding density is 3.35 gcm$^{-3}$.
This is slightly higher than the density of the WWW network, but still 
less than diamond's by $\sim$ 3\%. We conclude that this is the upper 
limit in the possible densities of ta-C. The highest densities for ta-C 
reported until now by experiment are less than 3.3 gcm$^{-3}$.

Unfortunately, experimental results do not provide a
clear picture of this issue. Different growth and characterization
techniques give sets of data which show a linear variation within
the respective set, but not when viewed all together. (A thorough discussion
of this point is given in Ref.\ \cite{Mathiou04}.) Very good agreement 
between theory and experiment holds for the ta-C region, i.e., for
densities higher than $\sim$ 2.8 gcm$^{-3}$. At lower densities,
experimental points scatter from method to method. For example,
data extracted from samples prepared by filtered
cathodic vacuum arc (FCVA) deposition \cite{Fallon,FerrPRB00}
differ from data extracted from samples prepared by
magnetron sputtering (MS) \cite{Schwan97}. 
A linear fit over the FCVA data was carried out 
by Ferrari {\it et al.} \cite{FerrPRB00}, yielding
$\rho$ (g/cm$^{3})$ = 1.92 + 1.37 ($sp^{3}$ fraction). This gives a
density of $\sim$ 3.3 gcm$^{-3}$ for 100\% $sp^{3}$ content, in good
agreement with our upper limiting value, but the lower limit at 
1.92 gcm$^{-3}$ is higher than ours, suggesting that $sp^{3}$ hybrids 
are absent in networks with lower densities. This can not explain reports 
of $sp^{3}$ sites in low-density carbyne films \cite{Barborini,Ravagnan}.
Note, however, that there are uncertainties in the measurements, usually 
by EELS, of the $sp^{3}$ content in such films.

An equally interesting physical trend in a-C is the
variation of elastic moduli as a function of mean coordination.
We seek to find simple formulas able to predict the 
hardness and related properties for any given network, over the entire
range of densities.

Thorpe and collaborators \cite{Thorpe85,Djord2} suggested that
the elastic moduli of bond-depleted crystalline diamond lattices
and of bond-depleted ``amorphous diamond'' networks (WWW model)
follow a power-law behavior $c \sim (\bar{z} - \bar{z}_{f})^{\nu}$,
with the exponent taking the value 1.5 $\pm$ 0.2. This mean-field
equation is characteristic of percolation theory, and describes
the contributions to rigidity from the local components of the
system as they connect to each other. The critical
coordination $\bar{z}_f$ = 2.4, denotes the transition from rigid to
floppy behavior, and comes out of the constraint-counting model of
Phillips \cite{Philips79} and Thorpe \cite{Thorpe83}.

We examined whether more realistic a-C networks can be described 
by the constraint-counting model, and if their moduli exhibit a power-law
behavior. For this, we used the cells generated by TBMD simulations
and the EDTB model. As a representative quantity, we calculated
the equilibrium bulk modulus $B_{eq}$. The results for $B_{eq}$ for
several networks as a function of $\bar{z}$ are given in Fig. 3.
Also included in this figure is the computed $B_{eq}$ for diamond (428 GPa) 
and for the WWW model (361 GPa). The latter value coincides with
that calculated with the Tersoff potential for 
WWW \cite{Kel00,Kel94,Kel01Diam}. 
The computed data can be fitted to the power-law relation
\begin{equation}
B_{eq} = B_{0}\ \left(\frac{\bar{z} - \bar{z}_{f}}{\bar{z}_{0} - 
\bar{z}_{f}}\right)^{\nu},
\label{modul1}
\end{equation}
where $B_{0}$ is the bulk modulus of the fully tetrahedral 
amorphous network, for which $\bar{z}_{0}$ = 4.0. Letting all 
fitting parameters in Eq. (\ref{modul1}) free, we obtain $B_{0}$ = 361 GPa, 
which is exactly the computed value for WWW, $\bar{z}_f$ = 2.25, 
and $\nu$ = 1.6. (For a measure of the quality of the fit: 
$R^2$ = 0.9907). If we fix $\nu$ to be 1.5 ($R^2$ = 0.9906), 
we get 2.33 for $\bar{z}_f$, and if we fix $\bar{z}_f$ to be 2.4 
($R^2$ = 0.9904), we get 1.4 for $\nu$. We thus conclude that the 
variation confirms the constraint-counting theory of Phillips and Thorpe, 
with a critical coordination close to 2.4, and it has a power-law behavior 
with a scaling exponent $\nu = 1.5 \pm 0.1$. For convenience, let us use
$\nu = 1.5$, so the modulus obeys the relation
\begin{equation}
B_{eq} = 167.3\ (\bar{z} - 2.33)^{1.5}.
\label{modul2}
\end{equation}
Our theory also predicts that ``amorphous diamond'' is softer
than diamond by $\sim$ 10\%.

Comparison of these results with experimental moduli derived from 
surface acoustic waves \cite{Schultrich} (SAW) and surface Brillouin 
scattering \cite{FerrAPL99} (SBS) measurements is very good,
especially in the ta-C region. For example, the computed modulus for 
$\bar{z} \simeq$ 3.9 equals $\sim$ 330 GPa and nearly
coincides with the SBS data. The agreement is less good at lower 
coordinations, where a fit to experimental 
points \cite{Robertson02,Mathiou04} extrapolates to $\bar{z}_f$ = 2.6, 
higher than the constraint-counting prediction.

\subsection{Diamond nanocomposite films}

Diamond nanocomposites consist of diamond nanocrystals embedded in
an a-C matrix \cite{Lifshitz}. They are produced by chemical
vapor deposition (CVD) via a multistage process \cite{Lifshitz},
and they differ from pure nanodiamond films with no a-C
component \cite{Gruen}. All diamond nanocomposite films reported
until now contain a hydrogenated a-C matrix. Recently, nanodiamonds
in pure a-C have been successfully grown \cite{Shay}.
Their structure, either with or without H, is rather well known
experimentally, but their stability and most of their properties,
including mechanical, are not yet understood.

A first step towards a theoretical description of these films
was done recently in our group \cite{Fyta03}. We summarize here the
most important findings of this investigation, based on MC 
simulations with the Tersoff potential, and we also provide
supplementary new results from TBMD simulations using the EDTB model.

A representative diamond nanocomposite network, generated by TBMD, is
portrayed in Fig. 4. It shows a spherical diamond nanocrystal, 
whose diameter is 12.5 \AA, positioned in the
middle of the cell and surrounded by the a-C matrix. Part of the
image cells, due to the periodic boundary conditions, are also shown.
This would correspond to an ideal case with a homogeneous dense dispersion
of crystallites of equal size in the matrix, at regularly ordered
positions. The nanodiamond volume fraction is 31\%. The density of the
a-C matrix $\rho_{am}$ is 3 gcm$^{-3}$ and its mean coordination 
$\bar{z}_{am}$ is 3.8. The size of the diamond crystallite is smaller 
than seen experimentally, but the overall structure captures the 
essential features of CVD grown nanocomposite films, especially the
non-hydrogenated ones.

A crucial issue is the stability of the diamonds as a function of
the coordination/density of the embedding medium. This extensive 
investigation required the analysis of many composite structures
and it was done through MC simulations using larger
cells (4096 atoms) \cite{Fyta03}. The quantity of interest is the
formation energy of a nanocrystal $E_{form}$, which can describe
the interaction of the embedded configuration with the host. It is
defined as 
\begin{equation}
E_{form} = E_{total} - N_{a}E_{a} - N_{c}E_{c},
\label{form}
\end{equation}
where $E_{total}$ is the total cohesive energy of the composite system
(amorphous matrix plus nanocrystal), calculated directly from the
simulation, $E_{c}$ is the cohesive energy per atom of the respective 
crystalline phase, $N_{c}$ is the number
of atoms in the nanocrystal, $N_{a}$ is the number of atoms in the
amorphous matrix, and $E_{a}$ is the cohesive energy per atom of the
pure, undistorted amorphous phase (without the nanocrystal) with
coordination $\bar{z}_{am}$. A negative value of $E_{form}$ denotes stability
of the nanostructure, a positive value indicates metastability or
instability.

The variation of $E_{form}$ as a function of $\bar{z}_{am}$ for a diamond 
with a fixed size embedded in several matrices is shown in Fig. 5. 
$E_{a}$ was computed from a series of calculations on pure a-C cells.
(For details see Ref.\ \cite{Fyta03}.) The most striking result of this
analysis is that diamonds are stable in matrices with $\bar{z}_{am}$ higher
than 3.6 ($\rho_{am} \simeq$ 2.6 gcm$^{-3}$), and unstable, or metastable
depending on temperature, in matrices with lower densities.
This nicely explains experimental results from different laboratories
indicating that diamond nanocrystals precipitate in a dense a-C
matrix \cite{Lifshitz,Shay}.

One way of checking the stability of nanocrystals is to subject them to 
thermal annealing. A stable structure should be sustained in the 
amorphous matrix, while a metastable structure should shrink in favor of 
the host. Indeed, analysis of the structure of diamonds annealed at high
$T$ (1500 - 2000 K) reveals \cite{Fyta03} that metastable nanocrystals
become heavily deformed in the outer regions near the interface 
with the amorphous matrix. Since only a small core remains intact, this
means that the diamonds extensively shrink. On the other hand,
the stable nanodiamonds are only slightly deformed and retain their 
tetrahedral geometry.

In addition, a stable nanodiamond has, in principle, the potential to 
expand against the surrounding matrix, provided that the barriers for 
this transformation can be overcome, possibly by further annealing
or ion irradiation. This means that nucleation
of diamond cores in a dense matrix might lead, under the appropriate 
experimental conditions, to a fully developed nanostructured
material with large grains.

The observation that diamonds are stable only in dense matrices
suggests a quantitative definition of ta-C, vaguely referred to 
as the form of a-C with a high fraction of sp$^3$ bonding. 
We can define ta-C as the form of a-C with a fraction of sp$^3$ sites 
above 60\%, in which diamond nanocrystals are stable (see Fig. 5). 
In other words, the predominantly tetrahedral amorphous network of ta-C 
is able to sustain crystalline inclusions. Networks with sp$^3$ fractions
below 60\% do not belong to the class of ta-C materials, because they
can not be transformed into a stable nanocrystalline state.

The intrinsic stress of the diamond nanocrystals and of the whole
composite material is a crucial quantity. As for pure ta-C, the
average stress influences the adhesion properties of the films.
The stress within the nanodiamonds is indicative of their stability 
in a-C matrices. To examine these issues, we calculated
the stress fields in the nanocomposite cells, using as a probe the
tool of atomic level stresses, as in the case of single-phase amorphous
carbon \cite{Kel00,Kel94,Kel01Diam}. This gives us the ability to
extract the stress built up in the nanodiamond and separate it from
the stress in the matrix, by summing up the atomic stresses over the
desired region. 

The first important aspect of this analysis is that,
in all cases studied, the average intrinsic stress
in the fully relaxed composite material is less than 1 GPa, practically
zero, even in the highly tetrahedral cases. This means, as in the case of
pure ta-C, that diamond nanocomposite films are able to eliminate any 
compressive stress generated during growth, when brought into their 
equilibrium state, perhaps by moderate thermal annealing. 
The stress in as-grown films has not yet been experimentally 
reported.

The other important finding is that the stress in the
nanocrystal is always found to be tensile, while it is compressive
in the matrix, yielding a net zero stress. This contrast can be
explained by noting that the density of the embedding medium is lower
than that of the diamond inclusions. As a result, atoms in the latter
are forced to strech their bonds in order to conform with the lower
density of the environment. 

A typical example of the stress state and its variation in a nanodiamond 
is shown in Fig. 6. The nanocrystal has a diameter of 17 \AA, and is 
embedded in a ta-C matrix with $\bar{z}_{am}$ = 3.9. 
The atomic stresses are averaged over spherical shells starting from 
the center and moving towards the interface. Negative values denote 
tensile stresses \cite{Kel94}. The stress in the core
of the diamond is very small, since the effect of the medium is weak, 
but it rises up as we move outwards, especially near the interface.
This is logical. Atoms near the interface strongly feel the
influence of the medium. However, the average tensile stress is low, 
$\sim$ -1.5 GPa/atom, because the density gradient between the nanodiamond 
and the matrix is small.

Obviously, the larger the density gradient the higher the tension 
felt by the inclusion. 
For example, when $\bar{z}_{am}$ = 3.84, the nanodiamond stress is 
-6.3 GPa/atom, and when $\bar{z}_{am}$ = 3.75, it becomes -9 GPa/atom. 
This trend explains the lowering of the relative 
stability of diamonds as $\bar{z}_{am}$ gets 
smaller (Fig. 5). Tension substantially increases at the outer regions 
of the nanodiamond, leading to deformation and eventually to amorphization 
and shrinking. This is remarkably evident for nanodiamonds in the region 
of metastability, where the intrinsic tensile stress becomes huge. 
For example, for a matrix with $\bar{z}_{am}$ = 3.3, the average stress 
in a typical nanodiamond is -30 GPa/atom.

Finally, we briefly comment on the hardness of these nanocomposite
materials. Their mechanical properties are not yet measured 
experimentally. We have preliminary results of calculations of the
bulk modulus of diamond nanocomposites, generated by TBMD/EDTB
simulations, a typical example of which is shown in Fig. 4.
We find moduli which are considerably higher than moduli of single-phase 
films of the same density, as calculated using Eq. (\ref{modul2}).
For example, for a nanocomposite with a total $\bar{z} \simeq$ 3.75 
and a density of $\sim$ 2.85 gcm$^{-3}$, the modulus approaches 350 GPa.
Eq. (\ref{modul2}) predicts $\sim$ 280 GPa for a pure a-C network
having the same $\bar{z}$. This represents a drastic 25\% increase
in strength, and opens up the possibility for even harder ta-C
materials for coatings and MEMS applications. A comprehensive account 
of the elastic properties of diamond nanocomposites will be given elsewhere.

\section{CONCLUSIONS}

Results from TBMD and MC simulations of pure a-C and diamond nanocomposite 
networks are presented. Definite trends in a-C regarding
the variation of the $sp^3$ fraction and the bulk moduli as a function 
of coordination/density are shown to firmly hold. Nanodiamonds are
stable only in dense ta-C matrices. The nanocomposite films are harder 
than pure a-C films of the same density. They possess zero intrinsic
stress when they are fully relaxed.

%\newpage 
\begin{center}

\begin{figure}
\vspace*{2cm}
\includegraphics[width=0.4\textwidth]{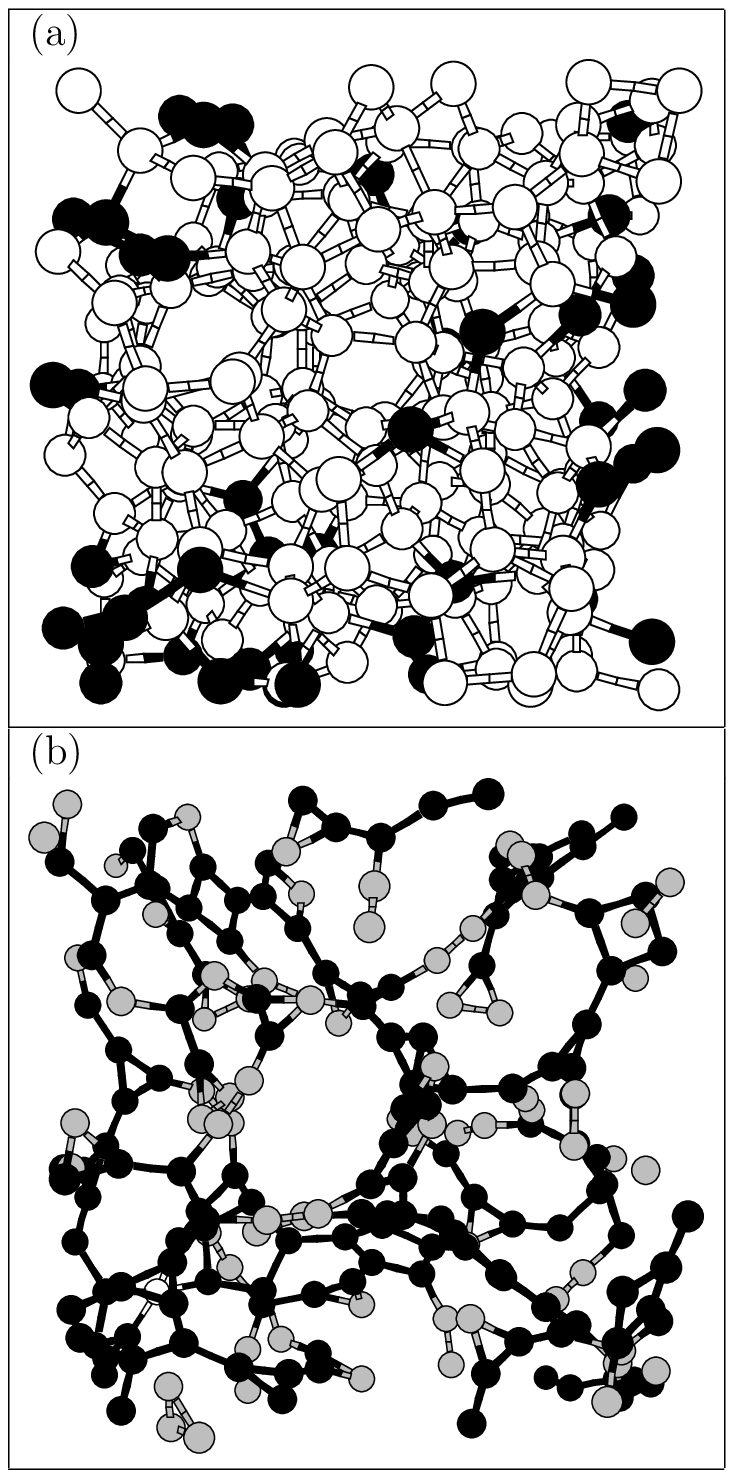}
\caption{Ball and stick models of representative a-C structures.
(a) A ta-C network with a density of 2.99 gcm$^{-3}$ and 79\%
$sp^3$ sites. (b) A low-density network (1.2 gcm$^{-3}$) with
66\% $sp^2$ sites and 33\% $sp^1$ sites. Open, dark, and shaded
spheres denote $sp^3$, $sp^2$, and $sp^1$ sites, respectively.}
\end{figure}

\begin{figure}
\includegraphics[width=0.4\textwidth,angle=270]{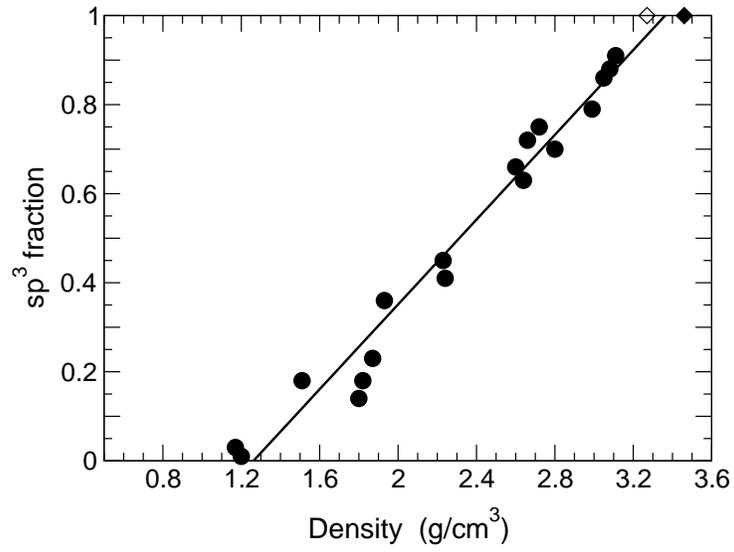}
\caption{The variation of $sp^3$ fraction as a function of density 
in a-C networks (filled circles). Line is a linear 
fit to the points. Also given are the corresponding calculated quantities 
for the WWW model (open diamond) and diamond (filled diamond).}
\end{figure} 

\begin{figure}
\vspace*{0.7cm}
\includegraphics[width=0.4\textwidth,angle=270]{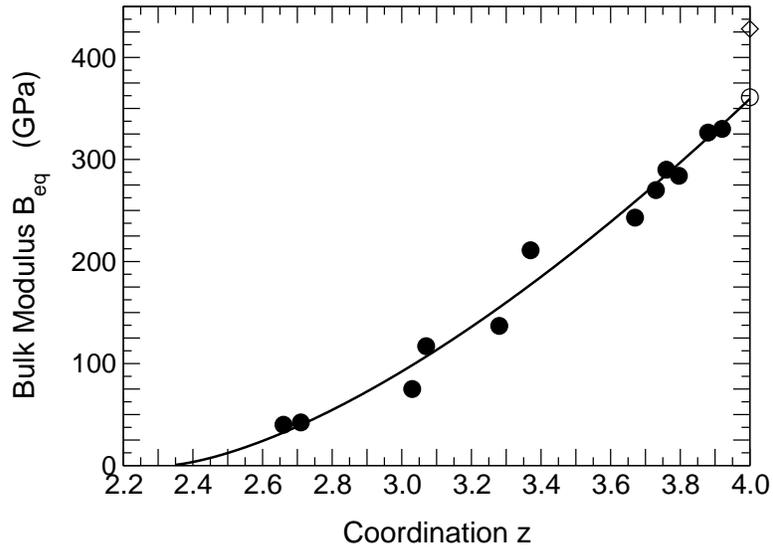}
\caption{The variation of computed bulk moduli (filled circles) of a-C 
networks as a function of the mean coordination $\bar{z}$. The open 
circle stands for the WWW model and the open diamond stands
for diamond.}
\end{figure} 

\begin{figure}
\includegraphics[width=0.45\textwidth]{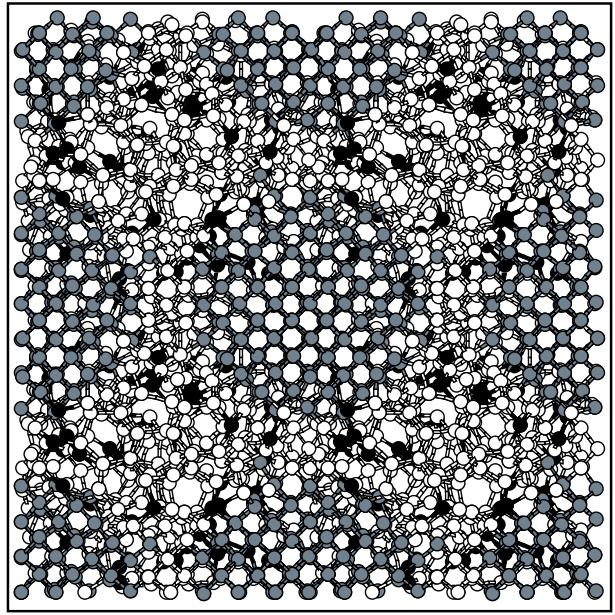}
\caption{Ball and stick model of a diamond nanocomposite network. The
nanodiamond with a diameter of 12.5 \AA \, is positioned in the center. 
Part of its neighboring images are also shown. The embedding a-C matrix
has a density of 3 gcm$^{-3}$. Shaded spheres denote atoms in the 
nanodiamond. Open (filled) spheres show $sp^3$ ($sp^2$) atoms in the
matrix.}
\end{figure} 

\begin{figure}
\includegraphics[width=0.5\textwidth]{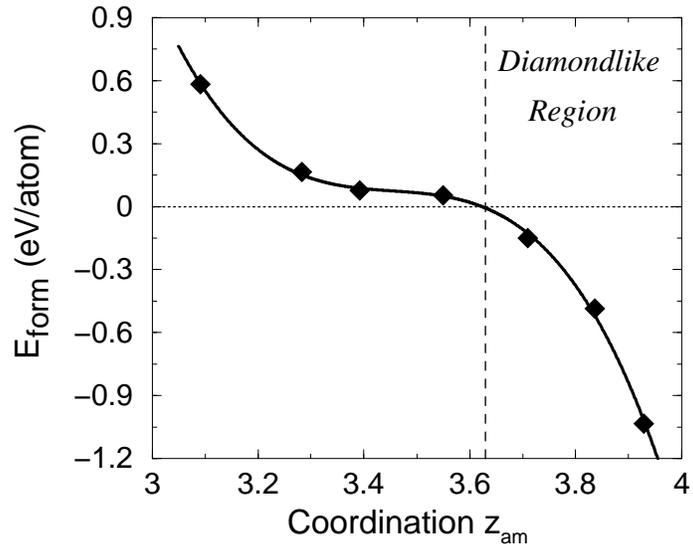}
\caption{Formation energies of a nanodiamond, having a
diameter of 18 \AA, embedded in several a-C matrices of different
mean coordination.}
\end{figure} 

\begin{figure}
\includegraphics[width=0.4\textwidth]{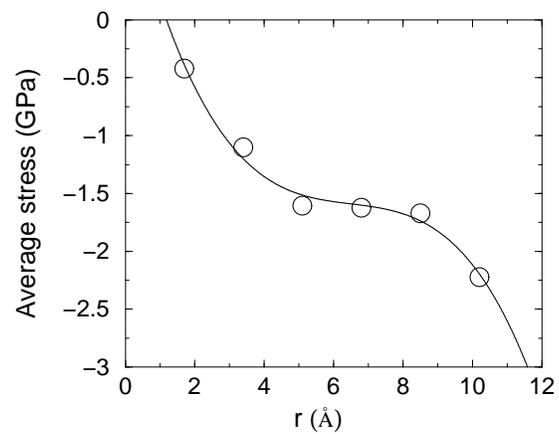}
\caption{Atomic stresses in a diamond nanocrystal (having a
diameter of 18 \AA) averaged over subsequent spherical shells of width
1.8 \AA, as a function of the distance from the center. The stresses
are calculated at 300 K. Line is a fit to the points.}
\end{figure} 
\end{center}
\end{document}